\documentclass[aps,eqsecnum,preprint,floats,epsf,epsfig,nofootinbib,showpacs]{revtex4}
\usepackage{graphicx}
\def\be{\begin{eqnarray}}
\def\en{\end{eqnarray}}
\def\non{\nonumber}

\def\bi{\bibitem}

\begin{document}

\title{\Large \bf Two-photon and two-gluon decays of $p$-wave heavy
quarkonium using a covariant light-front approach
 }

\author{ \bf  Chien-Wen Hwang\footnote{
t2732@nknucc.nknu.edu.tw} and Rurng-Sheng Guo\footnote{
t1953@nknucc.nknu.edu.tw}}

\affiliation{\centerline{Department of Physics, National Kaohsiung Normal University,} \\
\centerline{Kaohsiung, Taiwan 824, Republic of China}
 }


\begin{abstract}
In this paper, a study of two-photon and two-gluon decays in the
context of $p$-wave heavy quarkonia is presented. Within the
covariant light-front framework, the annihilation rates of scalar
and tensor quarkonium states are derived. In the absence of free
parameters in this case, the results for the charmonium decay widths
are consistent with the experimental data. However, in comparison to
other theoretical calculations, there are large discrepancies in our
results regarding bottomonia.

\end{abstract}
\pacs{14.40.Pq, 12.39.Ki}
\maketitle %

\section{Introduction}
Heavy quarkonium provides for a unique laboratory to study quantum
chromodynamics (QCD) regarding the bound states of a heavy
quark-antiquark system. Notably, the two-photon and two-gluon
annihilation rates of $p$-wave heavy quarkonium are helpful for
better understanding the details of quark-antiquark interaction and
can function as stringent tests for a potential model. Regarding
experimentation, the two-photon decay width of $\chi_{cJ}$ has been
measured by many laboratories \cite{PDG08,CLEO} and a new CLEO
measurement was reported recently \cite{CLEO}. Regarding theory,
relevant decay rates were first obtained through nonrelativistic
approximation \cite{NR1,NR2}; the relativistic corrections were
included within the Bethe-Salpeter equation
\cite{Munz,RSM1,RSM2,Laverty}, potential model \cite{Gupta},
relativistic quark model \cite{Godfrey,ebert1,ebert2},
nonrelativistic QCD factorization framework \cite{Schuler}, two-body
Dirac equations of constraint dynamics \cite{Crater}, effective
Lagrangian \cite{Lansberg}, and bound state perturbation theory
\cite{Lakhina}. The lattice calculation \cite{Dudek} and rigorous
QCD predictions \cite{Bodwin,Huang} were also applied. In addition,
the $s$-wave and $p$-wave electromagnetic and light hadronic
quarkonium decays in the heavy-quark velocity expansion were
computed in the nonrelativistic QCD approach \cite{BLS,BMV,BVM}.

This paper is aimed at the study of the two-photon and two-gluon
decay widths of $p$-wave heavy quarkonium states including the
scalar ($\chi_{c0}, \chi_{b0}, \chi'_{b0}$) and tensor
($\chi_{c2}, \chi_{b2}, \chi'_{b2}$) mesons.  
It is known that heavy quarkonium is relevant to nonrelativistic
treatments \cite{NR1,NR2,QR}. Although nonrelativistic QCD is a
powerful theoretical tool used to separate high-energy modes from
low-energy contributions, in most cases those attempting the
calculation of low-energy hadronic matrix elements have relied on
model-dependent nonperturbative methods. The light-front quark model
(LFQM) \cite{LFQM,LFQM1} is a relativistic quark model in which a
consistent and fully relativistic treatment of quark spins and the
center-of-mass motion can be carried out. This model has many
advantages. For example, the light-front wave function is manifestly
Lorentz invariant as it is expressed in terms of the momentum
fraction variables in analog to the parton distributions in the
infinite momentum frame. Moreover, hadron spin can also be correctly
constructed using the so-called Melosh rotation \cite{Melosh}. This
model is very well suited for studying hadronic form factors.
Specifically, as the recoil momentum increases (corresponding to a
decreasing $q^2$), we have to start seriously considering
relativistic effects. In particular, at the maximum recoil point
$q^2 = 0$ where the final-state particle could be highly
relativistic, there is no reason to expect that the nonrelativistic
quark model is still applicable.

The LFQM has been employed to obtain some physical quantities
\cite{Jaus0,Jaus1,JCC,CCH1}. However, one often calculates a
particular component (the ¡§plus¡¨ component) of the associated
current matrix element in the LFQM formulation. Because of the lack
of relativistic covariance, the results may show some
inconsistencies. The usual strategy of taking only the plus
component of the current matrix elements will ignore the zero-mode
contributions and render the matrix element noncovariant. As a
consequence, it is desirable to construct a covariant light-front
model that can provide a systematic way of exploring zero-mode
effects. Such a covariant model has been constructed in \cite{CCHZ}
for heavy mesons within the framework of heavy-quark effective
theory. Without appealing to the heavy-quark limit, a covariant
approach of the light-front model has been put forward for the usual
$s$-wave mesons \cite{Jaus2}, extended to the $p$-wave mesons
\cite{CCH2}, and employed in the context of the $s$-wave heavy
quarkonium \cite{Wei}. In this study, the $p$-wave heavy quarkonium
is explored through this covariant model. The details and formalism
are displayed in the next section.

The remainder of this paper is organized as follows. In Sec. II, the
formalism of a covariant light-front model is shown in cases of
scalar and tensor quarkonia and the annihilation rates of these
$p$-wave heavy quarkonium are derived. In Sec. III, after fixing the
parameters which appear in the trial wave function, the numerical
results and discussions are presented. Finally, conclusions are
given in Sec. IV.

\section{FORMALISM OF A COVARIANT LIGHT-FRONT MODEL }
\subsection{Formalism}
In the conventional light-front framework, the constituent quarks of
the meson are required to be on their mass shells (see Appendix A of
Ref. \cite{CCH2} for an introduction). 
Jaus \cite{Jaus2} proposed a covariant light-front approach that
permits a systematic way of dealing with zero-mode contributions.
Physical quantities can be calculated in terms of Feynman momentum
loop integrals which are manifestly covariant; this means that the
constituent quarks of the bound state are off shell. In principle,
this covariant approach will be useful if the vertex functions can
be determined by solving the QCD bound state equation. In practice,
we have to be content with phenomenological vertex functions, such
as those employed in the conventional light-front model. Therefore,
by using the light-front decomposition of the Feynman loop momentum,
say $p_\mu$, and integrating out the minus component of the loop
momentum $p^-$, one moves from the covariant calculation to the
light-front one. Moreover, the spectator quark is forced to be on
its mass shell after $p^-$ integration. Consequently, the covariant
vertex functions can be replaced by the phenomenological light-front
ones.

As stated in passing, in going from the manifestly covariant Feynman
integral to the light-front integral, the latter is no longer
covariant as it can receive additional spurious contributions
proportional to the lightlike four vector
$\omega^\mu=(\omega^-,\omega^+,\omega_\bot)=(2,0,0\bot)$. The
undesired spurious contributions can be eliminated by the inclusion
of the zero-mode contribution which amounts to performing the $p^-$
integration. The advantage of this covariant light-front framework
is that it allows a systematic way of handling the zero-mode
contributions and permits the user to obtain covariant matrix
elements.

To begin with, we consider the decay amplitudes given by one-loop
diagrams, as shown in Fig. 1 for the two-photon decay of $p$- wave
quarkonium states.
\begin{figure}[htbp]
\includegraphics*[width=5in]{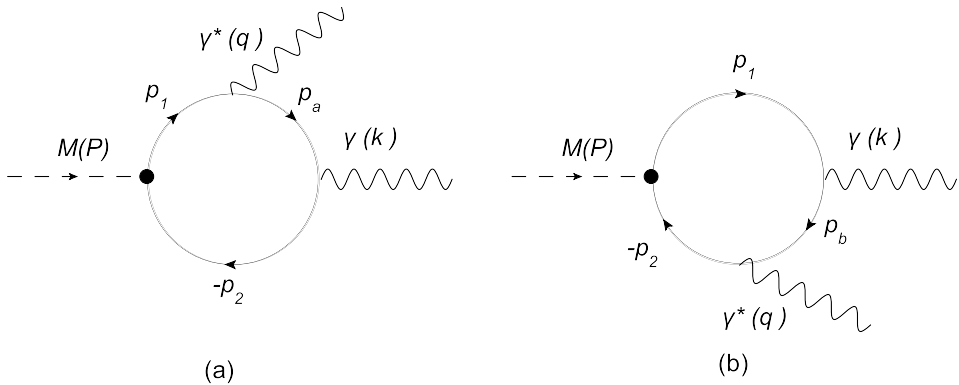}
\caption{Feynman diagrams for the $M\to\gamma\gamma^*$ process where
$P$ in the parentheses denotes the momentum of the meson. (b) is
related to (a) by the exchange of two photons.}
 \label{fig:1}
\end{figure}
The incoming meson has the momentum $P = p_1 + p_2$, where $p_1$ and
$p_2$ are the momenta of the off-shell quark and antiquark,
respectively, with masses $m$. These momenta can be expressed in
terms of the internal variables $(x_i, p_\perp)$,
 \be
 p_{1,2}^+=x_{1,2} P^+,\qquad p_{1,2\bot}=x_{1,2} P_\bot \pm p_\bot,
 \en
with $x_1+x_2 =1$. Note that we use $P = (P^+, P^-, P_\bot)$, where
$P^\pm = P^0 \pm P^3$, so that $P^2 = P^+P^--P^2_\bot$. In the
covariant light-front approach, total four momentum is conserved at
each vertex where quarks and antiquarks are off shell. However, it
is useful to define some internal quantities analogous to those
defined for on-shell quarks:
 \be
 e=\sqrt{m^2+p^2_\bot+p^2_z}, \quad p_z=\frac{(x_2-x_1)
 M_0}{2},\quad M^2_0 = (2 e)^2 =\frac{p^2_\bot+m^2}{x_1 x_2}.
 \en
where $m$ is the mass of heavy quark $c$ or $b$. Here, $M_0^2$ can
be interpreted as the kinetic invariant mass squared of the incoming
$q\bar q$ system and $e_i$ the energy of the quark $i$.

We need Feynman rules for the meson-quark-antiquark vertices to
calculate the amplitudes shown in Fig. 1. These Feynman rules for
vertices of scalar ($S$) and tensor ($T$) mesons are
 \be
 i\Gamma_M =\Bigg\{\begin{array}{l}
            -iH_S, {\rm~~~~~~~~~~~~~~~~~~~~~~~~~~~~~~~~~~~~~~~~~for}~ ^3 P_0, \\
            \frac{i}{2}H_T[\gamma_\mu-\frac{1}{W_V}(p_1-p_2)_\mu]
            (p_1-p_2)_\nu, {\rm~~~~for}~ ^3
            P_2.
            \end{array} \label{vertices}
 \en
It has been shown in Ref. \cite{12} that one can pass to the
light-front approach by integrating out the $p^-$ component of the
internal momentum in covariant Feynman momentum loop integrals. 
The specific form of the covariant vertex functions for on-shell
quarks can be determined through a comparison with the conventional
vertex functions as shown in Appendix A of Ref. \cite{CCH2}. They
are obtained as
 \be
 H_M &\to& H_M(\hat{p}_1^2,\hat{p}_2^2) \equiv h_M,\non \\
 W_V &\to& W_V(\hat{p}_1^2,\hat{p}_2^2) \equiv w_V,
 \en
where $\hat{p}_1=P-\hat{p}_2$ and $\hat{p}_2^2=m^2$. The form of the
function $h_M$  contains two parts: one is the momentum distribution
amplitude $\phi(x_i, p_\bot)$ which is the central ingredient in
light-front QCD, while the other is a spin wave function which
constructs a state of definite spin $(S, S_z)$ out of light-front
helicity eigenstates $(\lambda_1, \lambda_2)$. The spin wave
function is constructed by using the Melosh transformation
\cite{Melosh} and its spin structure is shown in Eq.
(\ref{vertices}). The explicit forms of $h_M$ and $w_V$ are given by
 \be
 h_S&=&(M^2-M_0^2)\sqrt{\frac{x_1x_2}{N_c}}\frac{1}{\sqrt{2}M_0}\frac{M_0}
 {2\sqrt{3}}~\phi,\non \\
 h_T&=&(M^2-M_0^2)\sqrt{\frac{x_1x_2}{N_c}}\frac{1}{\sqrt{2}M_0}~\phi,\non
 \\
 w_V&=&M_0+2 m.
 \en
The momentum distribution amplitude $\phi(x_i, p_\bot)$ is the
generalization of the distribution amplitude $\phi(x_i)$ of the pQCD
method and can be chosen to be normalizable, i.e., it satisfies
 \be
 \int \frac{dxd^2p_\bot}{2(2\pi)^3}|\phi(x,p_\bot)|^2 =
 1.\label{norm}
 \en
In principle, $\phi(x, p_\bot)$ is obtained by solving the
light-front QCD bound state equation $H_{LF} |\Psi\rangle =
M|\Psi\rangle$ which is the familiar Schr$\ddot{\rm o}$dinger
equation in ordinary quantum mechanics and $H_{LF}$ is the
light-front Hamiltonian. However, at the present time, methods
concerning how one can solve bound states equations are still
unknown. We are satisfied with utilizing some phenomenological
momentum distribution amplitudes which have been constructed
phenomenologically in describing hadrons. One widely used form is
the Gaussian type which we will employ in the application of the
covariant light-front approach.

\subsection{Two-photon decay widths of $p$-wave heavy quarkonium}
Quarkonia are eigenstates of the charge conjugation operator $C$
with eigenvalues $C=(-1)^{L+S}$, and the charge conservation
requires charge conjugation $C = +1$ state coupling to two real
photons. Thus, only $^3 P_0$ and $^3 P_2$ levels of the $p$-wave
quarkonium states can transform into two real photons. In the
process, $M\to \gamma\gamma$ , the final two photons are both on
shell. For the purpose of illustration, it is useful to consider a
more general process: $M\to \gamma\gamma^*$ with one photon off
shell. We introduce some transition form factors $f_i(q^2)$ arising
from the $M \gamma\gamma^*$ vertex. The $M\to \gamma\gamma$ process
is related to the form factors at $q^2=0$, i.e., $f_i(0)$. It is
worth mentioning that, however, the state $M$ in the process $M \to
\gamma\gamma^*$ is not only the $^3 P_0$ and $^3 P_2$ levels.

The matrix elements of the decay $S\to \gamma \gamma^*$ and $T\to
\gamma \gamma^*$ have the following structures \cite{ebert1}:
 \be
 {\cal A}^S_\mu&\equiv&\langle \gamma(k,\epsilon)| \bar q \gamma_\mu q
 |S(P)\rangle = f_0(q^2) \left[\epsilon_\mu (P\cdot k)-
 k_\mu (P\cdot \epsilon)\right],\\
 {\cal A}^T_\mu&\equiv&\langle \gamma(k,\epsilon)| \bar q \gamma_\mu q
 |T(P,\varepsilon)\rangle\non\\
 &=& f_1(q^2) \left[\epsilon_\mu (q \cdot \varepsilon \cdot k)-
 k_\mu (q \cdot \varepsilon \cdot \epsilon)+\varepsilon_{\mu\nu} \epsilon^\nu
 (q\cdot k)-\varepsilon_{\mu\nu} k^\nu (\epsilon\cdot q)\right]\non
 \\
 &+&f_2(q^2) \left[\epsilon_\mu (q\cdot k)-k_\mu (\epsilon
 \cdot q)\right]\frac{q\cdot \varepsilon \cdot k}{q\cdot k},
 \en
where $\epsilon$ is the polarization vector of the on-shell photon,
$q=P-k$ is the momentum transfer, and $\varepsilon$ is the
polarization of the tensor meson which satisfies the following
relations:
 \be
 &&P^{\mu}\varepsilon_{\mu\nu}(\lambda)=P^{\nu}\varepsilon_{\mu\nu}(\lambda)=0,~~~
 \varepsilon_{\mu\nu}(\lambda)=\varepsilon_{\nu\mu}(\lambda),~~~
 \varepsilon_{\mu\mu}(\lambda)=0,~~~
 \varepsilon^{\mu\nu}(\lambda)\varepsilon^*_{\mu\nu}(\lambda')=\delta_{\lambda\lambda'},\non\\
 &&\sum_{\lambda}\varepsilon_{\mu\nu}(\lambda)\varepsilon^*_{\alpha\beta}(\lambda)=\frac{1}{2}
 M_{\mu\alpha}M_{\nu\beta}+\frac{1}{2}
 M_{\mu\beta}M_{\nu\alpha}-\frac{1}{3}
 M_{\mu\nu}M_{\alpha\beta},
 \en
with $\lambda=\pm 2,\pm1,0$ representing the tensor meson helicities
and $M_{\mu\nu}=g_{\mu\nu}-P_\mu P_\nu/M^2$. There is no explicit
representation of the meson polarization tensor
$\varepsilon_{\mu\nu}(\lambda)$. The transition amplitude for the
processes of $S\to\gamma\gamma^*$ and $T\to \gamma \gamma^*$ can be
derived from the common Feynman rules and the vertices for the
meson-quark-antiquark coupling given in Eq. (\ref{vertices}). In the
covariant light-front approach, the meson is on shell while the
constituent quarks are off shell. To the lowest order approximation,
$S(T)\to\gamma\gamma^*$ is a one-loop diagram and is depicted in
Fig. 1. The amplitude is given as a momentum integral
 \be
 {\cal A}^S_\mu=ie^2_q e^2 N_c \int\frac{d^4
 p_1}{(2\pi)^4}&\Bigg\{&\frac{H_S}{N_1N_2N_a}{\rm Tr}\left[(-\not\! p_2+m)
 \not\!\epsilon(\not\! p_a+m)\gamma_\mu(\not\!p_1+m)\right]\non \\
 &+&\frac{H_S}{N_1N_2N_b}{\rm Tr}\left[(-\not\! p_2+m)
 \gamma_\mu(\not\!
 p_b+m)\not\!\epsilon(\not\!p_1+m)\right]\Bigg\},\label{AS}\\
 {\cal A}^T_\mu=ie^2_q e^2 N_c \int\frac{d^4
 p_1}{(2\pi)^4}&\Bigg\{&\frac{H_T}{N_1N_2N_a}{\rm Tr}\left[t(-\not\! p_2+m)
 \not\!\epsilon(\not\! p_a+m)\gamma_\mu(\not\!p_1+m)\right]\non \\
 &+&\frac{H_T}{N_1N_2N_b}{\rm Tr}\left[t(-\not\! p_2+m)
 \gamma_\mu(\not\!
 p_b+m)\not\!\epsilon(\not\!p_1+m)\right]\Bigg\},\label{AT}
 \en
where
 \be
 &&p_a=p_1-q,\qquad\qquad\quad p_b=q-p_2, \non \\
 &&N_1=p_1^2-m^2+i\epsilon,\qquad N_2=p^2_2-m^2+i\epsilon, \non \\
 &&N_a=p_a^2-m^2+i\epsilon,\qquad N_b=p^2_b-m^2+i\epsilon, \non \\
 &&t=\varepsilon_{\alpha\beta}
 \left[\gamma^\alpha-\frac{1}{W_V}(p_1-p_2)^\alpha\right]
 \frac{(p_1-p_2)^\beta}{2},
 \en
and $e_q$ is the electric charge of the quark: $e_q = 2/3$ for the
$c$ quark and $e_q = -1/3$ for the $b$ quark. The first and second
terms in Eq. (\ref{AS}) come from diagrams Figs. 1 (a) and 1 (b),
respectively. In the calculation, it is convenient to choose the
purely transverse frame $q^+ = 0$, i.e., $q^2 = -q^2_\bot \leq 0$.
The advantage of this choice is that there is no so-called $Z$-
diagram contributions. The sacrifice associated with this approach
is that only the form factor at spacelike regions can be calculated
directly. The values at the timelike momentum transfer $q^2 > 0$
regions are obtained through analytic continuation. In this study,
the continuation is not necessary because we only need the form
factors at $q^2 = 0$ for the $S(T)\to \gamma\gamma$ process.

First, we discuss the calculation of Fig. 1 (a). The factors $N_1$,
$N_2$, and $N_a$ produce three singularities in the $p^-_1$ complex
plane: one lies in the upper plane and the other two are in the
lower plane. By closing the contour in the upper $p^-_1$ complex
plane, the momentum integral can be easily calculated since there is
only one singularity in the plane. This corresponds to putting the
antiquark on the mass shell. Given this restriction, the momentum
$p_2 \to \hat{p}_2$ with $\hat{p}_2^2=m^2$, and $\hat{p}_1 =
P-\hat{p}_2$. The on-shell restriction and the requirement of
covariance lead to the following replacements:
 \be
 N_1 &\to& \hat{N}_1 = x_1(M^2-M^2_0),\non \\
 N_2 &\to& \hat{N}_2 = \hat{N}_1+(1-2x_1)M^2=x_2 M^2-x_1 M^2_0,\non \\
 N_a &\to& \hat{N}_a = x_2 q^2-x_1M^2_0+2 p_\bot \cdot q_\bot,\non \\
 \int\frac{d^4p_1}{(2\pi^4)}\frac{H_{S,T}}{N_1N_2N_a} &\to&
 -i\pi
 \int\frac{dx_2d^2p_\bot}{(2\pi^4)}\frac{h_{S,T}}{x_2\hat{N}_1\hat{N}_a}.
 \en
For Fig. 1 (b), the contour is closed in the lower $p^-_1$ complex
plane. It corresponds to putting the quark on the mass shell and the
momentum $p_1\to \hat{p}_1$ with $\hat{p}^2_1 = m^2$. For this
scenario, we need to do the following replacements:
 \be
 N_1 &\to& \hat{N}_1 = x_1 M^2-x_2 M^2_0x_1(M^2-M^2_0),\non \\
 N_2 &\to& \hat{N}_2 = x_2(M^2-M^2_0),\non \\
 N_b &\to& \hat{N}_a = x_1 q^2-x_2 M^2_0 - 2 p_\bot \cdot q_\bot,\non \\
 \int\frac{d^4p_1}{(2\pi^4)}\frac{H_{S,T}}{N_1N_2N_b} &\to&
 -i\pi
 \int\frac{dx_2d^2p_\bot}{(2\pi^4)}\frac{h_{S,T}}{x_1\hat{N}_2\hat{N}_b}.
 \en
After the above treatments, the transition amplitudes of $S \to
\gamma\gamma^*$ and $T \to \gamma\gamma^*$ for Fig. 1 (a), for
example, are obtained as
 \be
 {\cal A}_\mu^{S(a)}&=&N_c e_q^2e^2
 \int\frac{dx_2d^2p_\bot}{4\pi^3}\frac{h_S}{x_1 x_2 (M^2-M_0^2)}
 \frac{m}{-x_2q^2+x_1M_0^2-2 p_\bot\cdot q_\bot}\frac{M_0}{2\sqrt{3}}\non \\
 &\times& \big\{\epsilon_\mu(m^2- P\cdot q - p_1^2+2 p_1\cdot q)+2
 p_{1\mu}(2\epsilon\cdot p_1-\epsilon\cdot P-\epsilon\cdot q)+
 q_\mu(\epsilon\cdot P-2\epsilon\cdot p_1)\non \\
 &&+P_\mu\epsilon\cdot q\big\},\label{Sa}\\
 {\cal A}_\mu^{T(a)}&=&N_c e_q^2e^2
 \int\frac{dx_2d^2p_\bot}{4\pi^3}\frac{h_T}{x_1 x_2 (M^2-M_0^2)}
 \frac{1}{-x_2q^2+x_1M_0^2-2 p_\bot\cdot q_\bot}\non \\
 &\times& \bigg\{\epsilon_\mu\left[(m^2+P\cdot p_1-p_1^2) p_1\cdot\varepsilon\cdot q
 +(m^2-P\cdot q+2p_1\cdot q-p_1^2)\left(1-\frac{2 m}{w_V}\right)
 p_1\cdot\varepsilon\cdot p_1 \right]\non\\
 &&+P_\mu\left[(m^2+p_1\cdot q-p_1^2)\epsilon\cdot \varepsilon\cdot p_1-
 \epsilon\cdot p_1 p_1\cdot\varepsilon\cdot q+\epsilon\cdot q\left(1-\frac{2 m}{w_V}\right)
 p_1\cdot\varepsilon\cdot p_1\right]\non \\
 &&+q_\mu\left[(p^2_1-m^2-P\cdot p_1)\epsilon\cdot \varepsilon\cdot p_1+
 (\epsilon\cdot P-2 \epsilon\cdot p_1)\left(1-\frac{2 m}{w_V}\right)p_1\cdot\varepsilon
 \cdot p_1\right]\non \\
 &&+p_{1\mu}\bigg[(m^2-P\cdot q+2P\cdot p_1-p_1^2) \epsilon\cdot \varepsilon\cdot p_1
 +\epsilon\cdot P p_1\cdot\varepsilon\cdot q\non \\
 &&\qquad+(4\epsilon\cdot p_1-2\epsilon\cdot q-
 2\epsilon\cdot P)\left(1-\frac{2 m}{w_V}\right)p_1\cdot\varepsilon\cdot p_1 \bigg]\non \\
 &&+\varepsilon_{\mu\nu}p_1^\nu\bigg[m^2(\epsilon\cdot p_1-\epsilon\cdot P-\epsilon\cdot q)
 +(\epsilon\cdot p_1 P\cdot q-\epsilon\cdot P p_1\cdot q-\epsilon\cdot q P\cdot p_1)\non \\
 &&\qquad\quad+p_1^2(\epsilon\cdot q+\epsilon\cdot P-\epsilon\cdot p_1)\bigg]\bigg\}.\label{Ta}
 \en
The integration of $p_{1\mu}$, $p_{1\mu}p_{1\nu}$ ,
$p_{1\mu}p_{1\nu}p_{1\alpha}$, and
$p_{1\mu}p_{1\nu}p_{1\alpha}p_{1\beta}$ in Eqs. (\ref{Sa}) and
(\ref{Ta}) can be expressed in terms of three external vectors:
$\widetilde{P}$, $q$, and $\omega$, as in the Appendix. The
transition amplitude of $S(T) \to \gamma\gamma^*$ for Fig. 1 (b) can
be obtained through a similar process. If we choose the frame where
the meson is at rest and the photons travel in the $\pm z$
directions, then the two-photon decay amplitude of $S \to
\gamma\gamma$ is obtained as
 \be
 &&{\cal M}^S(S\to\gamma\gamma)=({\cal A}^{S(a)}_\mu+{\cal A}^{S(b)}_\mu)
  |_{q^2=0} \cdot \epsilon'\non \\
  &=&e^2_q e^2 \sqrt{\frac{N_c}{2}}\epsilon\cdot \epsilon'
  \int\frac{dx_2d^2
  p_\bot}{8\pi^3}\frac{m \phi}{\sqrt{m^2+p_\bot^2}}\left(\frac{2x_1
  M_0^2-M^2-4p^2_\bot}{2\sqrt{3}x_1 M_0}+\frac{2x_2
  M_0^2-M^2-4p^2_\bot}{2\sqrt{3}x_2 M_0}\right),\label{Mscalar}
 \en
where $\epsilon'$ is the polarization of another photon. The decay
rate of the process $S\to \gamma\gamma$ is \cite{choiphd}
 \be
 \Gamma(S\to \gamma\gamma)=\frac{s}{8\pi M_S^2}|\vec{k}|\sum_{{\rm pol}}|{\cal
 M}^S|^2,\label{dwSgg}
 \en
where $s=1/2$ for two identical photons and the photon momentum
$|\vec{k}|=M/2$. Regarding the case of $T \to \gamma\gamma$, in the
practical calculations below, it is convenient to represent
$\varepsilon_{\mu\nu}$ in the following form \cite{polartensor}:
 \be
 \varepsilon_{\mu\nu}(\lambda)&=&\frac{\sqrt{6}}{12}(2-|\lambda|)(1-|\lambda|)
 \left[3 n^\mu_3n^\nu_3+\left (g_{\mu\nu}-\frac{P_\mu
 P_\nu}{M^2}\right)\right]\non\\
 &&+\frac{1}{4}(1-|\lambda|)\left[\lambda(n^\mu_1n^\nu_1-n^\mu_2n^\nu_2)
 +i|\lambda|(n^\mu_1n^\nu_2+n^\mu_2n^\nu_1)\right]\non\\
 &&+\frac{1}{2}(2-|\lambda|)\left[\lambda(n^\mu_1n^\nu_3+n^\mu_3n^\nu_1)
 +i|\lambda|(n^\mu_2n^\nu_3+n^\mu_2n^\nu_1)\right],
 \en
where $n^\mu_i (i=1,2,3)$ is the basis vector satisfying $n^\mu_i
n^\nu_j g_{\mu\nu}=g_{ij}$. In the center-of-mass frame, we may have
$P^\mu=(E,0,0,|\textbf{P}|)$ and choose the basis with collinear
$n^\mu_3$ and $P^\mu$ vectors as the simplest one
 \be
 n^\mu_1=(0,1,0,0),~~~n^\mu_2=(0,0,1,0),~~~n^\mu_3=\frac{1}{M}(|\textbf{P}|,0,0,E).
 \en
For the frame in which the meson is at rest and the photons travel
in the $\pm z$ directions, the decay amplitudes of $T\to
\gamma\gamma$ are obtained for $ \lambda=0$ and $\pm 2$
 \be
 {\cal M}^T_{\lambda=0}(T\to \gamma\gamma)&=&({\cal A}^{T(a)}_\mu+{\cal
 A}^{T(b)}_\mu)|_{q^2=0,\lambda=0}\cdot \epsilon'\non \\
 &=&e^2_q e^2 \sqrt{\frac{N_c}{2}}\sqrt{\frac{2}{3}}\epsilon\cdot \epsilon'
  \int\frac{dx_2d^2
  p_\bot}{16\pi^3}\frac{\phi}{\sqrt{m^2+p_\bot^2}}\left(\frac{t_{0a}}{x_1 M_0^2}+\frac{t_{0b}}{x_2
  M_0^2}\right),\label{T0}\\
 {\cal M}^T_{\lambda=\pm 2}(T\to \gamma\gamma)&=&({\cal A}^{T(a)}_\mu+{\cal
 A}^{T(b)}_\mu)|_{q^2=0,\lambda=\pm 2}\cdot \epsilon'\non \\
 &=&e^2_q e^2 \sqrt{\frac{N_c}{2}}
  \int\frac{dx_2d^2
  p_\bot}{8\pi^3}\frac{\phi}{\sqrt{m^2+p_\bot^2}}\left(\frac{t_{2a}}{x_1 M_0^2}+\frac{t_{2b}}{x_2
  M_0^2}\right),\label{T2}
 \en
where
 \be
 t_{0a}&=&x_1(m^2-x_2 M^2)(x_1M_0^2 - x_2 M^2)+ p_\bot^2 [-3 m^2 + (1
 - 4 x_1 + 2 x_1^2)M^2]-p_\bot^4\non \\
 &+&\frac{m}{w_V} \left\{ x_1(2 m^2 - M^2) (x_2 M^2-x_1 M_0^2) +
   p^2_\bot [2 m^2 - 4 x_1^2 M_0^2 + ( 4 x_1
   x_2-1)M^2]-2p^4_\bot\right\},\non \\
 t_{0b}&=& t_{0a}|_{x_1\leftrightarrow x_2},\\
 t_{\pm 2a}&=&p^2_\bot \left(m^2+M_0^2 x_1^2+2 \frac{m}{w_V}p^2_\bot\right)
 (i \epsilon'\cdot n_2 \epsilon\cdot n_1
 +i\epsilon\cdot n_2 \epsilon'\cdot n_1\pm\epsilon'\cdot n_1 \epsilon\cdot n_1
 \mp \epsilon\cdot n_2\epsilon'\cdot n_2),\non \\
 t_{\pm 2b}&=& t_{2a}|_{x_1\leftrightarrow x_2}.
 \en
The derivations of Eqs. (\ref{T0}) and (\ref{T2}) use the formulas
 \be
 \int d^2 p_\bot \left[p_\bot \cdot q_\bot, (p_\bot \cdot q_\bot)^2, (p_\bot \cdot q_\bot)^3,
 (p_\bot \cdot q_\bot)^4 \right]=\int d^2 p_\bot\left[0,\frac{p^2_\bot q^2_\bot}{2},0,
 \frac{3 p^4_\bot q^4_\bot}{8}\right].
 \en
The amplitudes of $T\to \gamma\gamma$ for $\lambda=\pm1$ vanish
because the combination of the helicities of two final photons can
be $\pm 2$ or $0$, but never equal to $\pm1$. The decay rate of the
process $T\to \gamma\gamma$ is
 \be
 \Gamma(T\to \gamma\gamma)=\frac{s}{8\pi M_T^2}\frac{|\vec{k}|}{5}
 \sum_{\lambda=0,\pm1,\pm2}\sum_{{\rm pol}}|{\cal
 M}^T|^2. \label{dwTgg}
 \en
The factor $5$ in the denominator corresponds to $2J+1$, where $J$
is the total angular momentum of the meson.

Finally, the two-gluon decay width of quarkonium can be easily
obtained from the two-photon decay width, with a simple replacement
in the photon decay width formula
 \be
 e_q^4 \alpha^2 \to \frac{2}{9} \alpha_s^2.\label{replace}
 \en
\section{Numerical results and discussions}
In this section, the two-photon and two-gluon decay widths of
$p$-wave heavy quarkonium states are estimated. Prior to numerical
calculations, the parameters $m_{c,b}$ and $\beta_{c\bar c,b\bar
b}$, which appeared in the wave function, must be first determined.
We consider the Hamiltonian of the $s$- and $p$-wave heavy
quarkonium states as
 \be
 H_s&=&2\sqrt{m^2+\vec\kappa^2}+V_{\rm conf}-\frac{4\alpha_s}{3 r}+g_0 s_1 \cdot
 s_2,\label{Hs}\\
 H_p&=&2\sqrt{m^2+\vec\kappa^2}+V_{\rm conf}-\frac{4\alpha_s}{3 r}+g_1 S \cdot
 L+g_2 S_{12}+g_3 s_1 \cdot s_2,\label{Hp}
 \en
where $V_{\rm conf}=br(br^2)$ is the linear [harmonic
oscillator(HO)] potential, $S_{12}=(3 s_1 \cdot \hat r~s_2 \cdot
\hat r - s_1\cdot s_2)$ is the tensor force operator, and
$g_{0,1,2,3}$ are the functions of the relevant interquark
potentials (the details are shown in, for example, \cite{RR,BNS}).
In this way, not only is the spin-weighted average of the $s$-wave
states $M(S_J)\equiv[M(^1 S_0)+3 M(^3 S_1)]/4$ free of the spin-spin
contribution, but also the mass difference between the spin-single
ground state $M(^1 P_1)$ and the spin-weighted average of the
$p$-wave triplet states $M(^3 P_J)\equiv[M(^3 P_0)+3 M(^3 P_1)+5
M(^3 P_2)]/9$ has a contribution which comes from the spin-spin
interaction.\footnote{The calculations of the expectation values of
the fourth and fifth terms for the Hamiltonian Eq. (\ref{Hp}) can be
referred to in the appendix of Ref. \cite{Jackson}.} Experimentally
the latter hyperfine splitting is less than $1$ Mev \cite{PDG08} in
the charmonium sector and can be ignored here. Notably, we can use
the masses $M(S_J)$, $M(^3 P_J)$ and their variational principle for
the Hamiltonian equations. (\ref{Hs}) and (\ref{Hp}) to determine
parameters $m$ and $\beta$. In the process, the $1S$, $1P$, and $2P$
states harmonic wave functions
 \be
 \phi^{1S}(x,p_\bot)&=& 4\left(\frac{\pi}{\beta^2}\right)^{3/4}\sqrt{\frac{dp_z}{dx}}
 ~{\rm exp}\left[-\frac{|\vec{p}|^2}{2\beta^2}\right],\label{wfp1s}\\
 \phi^{1P}_m(x,p_\bot)&=& 4\left(\frac{\pi}{\beta^2}\right)^{3/4}
 \sqrt{\frac{dp_z}{dx}}\sqrt{\frac{2}{\beta^2}}
 p_m{\rm exp}\left[-\frac{|\vec{p}|^2}{2\beta^2}\right],\label{wfp1p}\\
 \phi^{2P}_m(x,p_\bot)&=& 4\left(\frac{\pi}{\beta^2}\right)^{3/4}
 \sqrt{\frac{dp_z}{dx}}\sqrt{\frac{5}{\beta^2}}
 p_m \left(1-\frac{2|\vec{p}|^2}{5 \beta^2}\right){\rm exp}\left[-\frac{|\vec{p}|^2}
 {2\beta^2}\right],\label{wfp2p}
 \en
which satisfy the normalization Eq. (\ref{norm}) and their conjugate
coordinate wave function
 \be
 \widetilde{\phi}^{1S}(r)&=& \left(\frac{\beta^2}{\pi}\right)^{3/4}
 ~{\rm exp}\left[-\frac{\beta^2 r^2}{2}\right],\label{wfr1s}\\
 \widetilde{\phi}^{1P}_m(r)&=& \sqrt{2}\left(\frac{\beta^2}{\pi}\right)^{3/4}~\beta r_m~{\rm exp}
 \left[-\frac{\beta^2 r^2}{2}\right],\label{wfr1p}\\
 \widetilde{\phi}^{2P}_m(r)&=& \sqrt{5}\left(\frac{\beta^2}{\pi}\right)^{3/4}
 \beta r_m \left(1-\frac{2\beta^2 r^2}{5}\right){\rm exp}
 \left[-\frac{\beta^2 r^2}{2}\right],\label{wfr2p}
 \en
where $a_{m=\pm1}=\mp(a_x\pm ia_y)/\sqrt{2}$ and $a_{m=0}=a_z$ are
needed. In addition to the coefficient of confined potential $b$,
the $c\bar c$ and $b \bar b$ sectors each have four parameters,
$m_q$, $\alpha_s$, $\beta^{1S}$, and $\beta^{1P}$, for the $1S$ and
$1P$ quarkonium states. Regarding the constraints, in addition to
the four masses $M_{c\bar c,b\bar b}(1S_J)$ and $M_{c\bar c,b\bar
b}(1^3 P_J)$, the four equations (\ref{wfp1s}), (\ref{wfr1s}) and
(\ref{wfp1p}), (\ref{wfr1p}) are used as the trial functions of the
variational principle for $1S$ and $1P$ states, respectively.
Therefore, $b$ is the only free parameter in our fitting. We employ
the data $\Gamma(\chi_{c0}\to\gamma\gamma)=2.36\pm0.35$ keV
\cite{PDG08}, to fix $b$; then the above parameters can be
determined and shown in Table 1.
\begin{table}
\caption{\label{tab:parameter} The relevant parameters $b$, $m_q$,
$\alpha_s$, and $\beta$ of the $p$-wave heavy quarkonium states. }
\begin{ruledtabular}
\begin{tabular}{c|crccc}
 Potential & $b$ & $m_q$ (GeV) & $\alpha_s$  & $\beta^{1P}$ (GeV) &  $\beta^{2P}$(
 GeV)\\\hline
Linear & $0.176^{+0.007}_{-0.008}$ GeV$^2$ & $c\bar
c$~~$1.42\mp0.02$ &
$0.489^{-0.018}_{+0.019}$ & $0.510\pm 0.002$ &\\
 & &  $b\bar b$~~$4.78\mp0.01$ & $0.399^{-0.006}_{+0.007}$ & $0.807 \pm 0.005$ & $0.568^{+0.025}_{-0.024}$\\\hline
 HO &$0.0490^{+0.0022}_{-0.0024}$ GeV$^3$ & $c\bar c$~~$1.42\mp 0.02$   &
$0.358^{-0.025}_{+0.018}$ & $0.577^{+0.001}_{-0.003}$&\\
 & &  $b\bar b$~~$4.87_{+0.00}^{-0.01}~~~$ & $0.425^{-0.005}_{+0.006}$ & $0.847 \pm 0.05$ & $0.571^{+0.007}_{-0.009}$
\end{tabular}
\end{ruledtabular}
\end{table}
For $2P$ states, due to the insufficient data regarding
$\chi'_{cJ}$, only the parameter $\beta^{2P}_{b\bar b}$ is
determined by the mass $M_{b\bar b}(2^3 P_J)$ and also revealed in
Table 1. There are three items in Table 1 worth mentioning: First,
the parameter $b=0.176^{+0.007}_{-0.008}$ GeV$^2$ in the linear
potential is consistent with the string tension $b=0.18$ GeV$^2$
which is well known from other quark model analyses
\cite{Godfrey,ISGW,SI}. Second, there are presently different
conceptions about the value of $\alpha_s$ in the low-energy region.
In lattice QCD \cite{latticeQCD} and the field correlator method
\cite{BNS}, for example, ones found for the coupling constant in the
static potential, parametrized as a linear plus Coulomb potential,
the small values $\alpha_s$ are $0.21$ and $0.16$, respectively. In
phenomenological potentials the Coulomb constant is larger:
$\alpha_s$ is $0.46$ \cite{ccPRL} and $0.43\pm 0.02$ \cite{bbPRD}
which corresponded to the charmonium and bottomonium states. Here
our values are consistent with the latter. Third, in the HO
potential, the strong coupling constant $\alpha_s$ of the $c\bar c$
sector is smaller than that of the $b\bar b$ sector, which violates
the concept of asymptotic freedom. The dependences of
$\alpha_s(c\bar c)$ and $\alpha_s({b\bar b})$ on $b$ regarding the
linear and HO potentials are shown in Figs. 2 and 3, respectively.
 \begin{figure}
 \includegraphics*[width=4in]{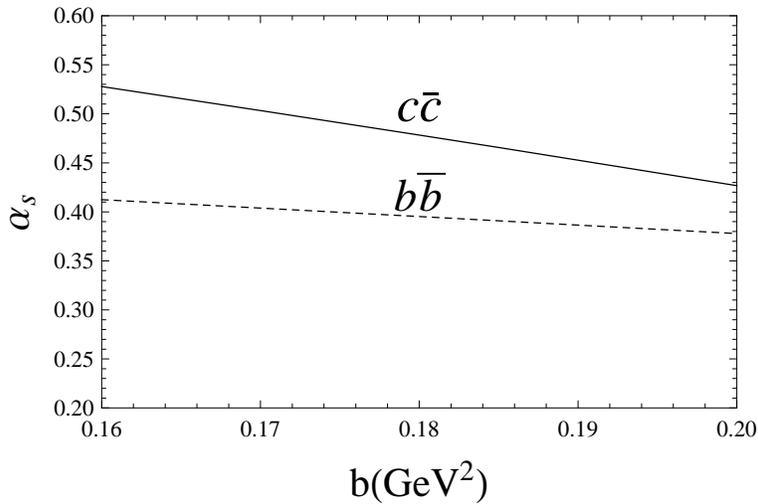}
 \caption{ Dependences of $\alpha_s(c\bar c)$ (solid
 line) and $\alpha_s({b\bar b})$ (dashed line) on $b$ in the linear potential}
  \label{fig:br1}
 \end{figure}
 \begin{figure}
 \includegraphics*[width=4in]{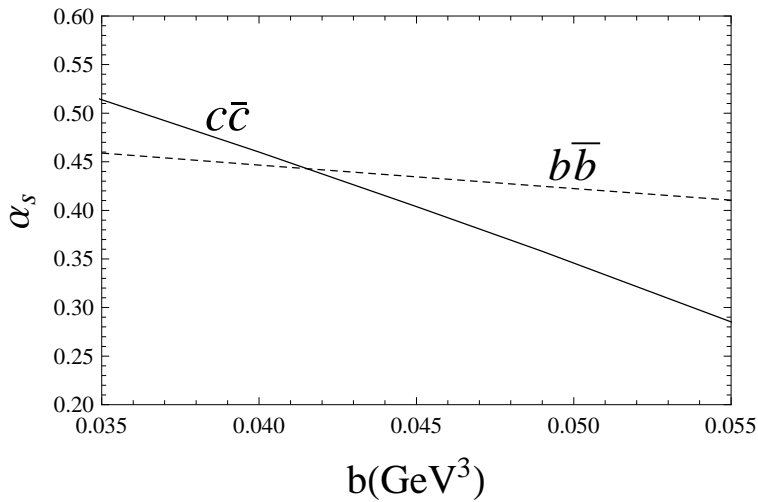}
 \caption{ Dependences of $\alpha_s(c\bar c)$ (solid
 line) and $\alpha_s({b\bar b})$ (dashed line) on $b$ in the HO potential.}
 \label{fig:br2}
 \end{figure}
Obviously, the HO potential is not suitable when $b\geq 0.0410$
GeV$^3$. However, in the data fitting of
$\Gamma(\chi_{c0}\to\gamma\gamma)$, we obtained $b=
0.0490^{+0.0022}_{-0.0024}$ GeV$^3$. Therefore, only the linear
potential is employed in the following calculations.

Next, we use the parameters of the linear potential in Table I and
Eqs. (\ref{dwSgg}), (\ref{dwTgg}), and (\ref{replace}) to calculate
the two-photon and two-gluon decay widths of the $\chi_{c0,c2}$,
$\chi_{b0,b2}$, and $\chi'_{b0,b2}$ states. The numerical results,
which compare with the experimental data and other theoretical
evaluations, are shown in Tables 2 and 3.
\begin{table}[ht!]
\caption{\label{tab:widthgaga} Two-photon widths of the $p$-wave
heavy quarkonium states. $R$ is defined as $\Gamma^{\chi_{c2}}_{2\gamma}/\Gamma^{\chi_{c0}}_{2\gamma}$.
}
\begin{ruledtabular}
{\footnotesize
\begin{tabular}{cccccccc}
  & $\Gamma^{\chi_{c0}}_{2\gamma}$ (keV) & $\Gamma^{\chi_{c2}}_{2\gamma}$ (keV)
  & R &
  $\Gamma^{\chi_{b0}}_{2\gamma}$ (eV)& $\Gamma^{\chi_{b2}}_{2\gamma} $ (eV)&
  $\Gamma^{\chi'_{b0}}_{2\gamma}$ (eV)& $\Gamma^{\chi'_{b2}}_{2\gamma}$
  (eV)\\\hline
  PDG \cite{PDG08} & $2.36\pm 0.35$ & $0.515\pm 0.060$& $0.218\pm 0.058$ & & & & \\
 This work & \underline{$2.36\pm 0.35$} & $0.346^{+0.009}_{-0.011} $&$0.147\pm0.026$& $8.02^{+1.12}_{-1.15}$ &
 $3.99^{+0.11}_{-0.13}$ & $2.26^{+0.16}_{-0.17}$ & $1.72^{+0.39}_{-0.33}$ \\
 M$\ddot{\rm u}$nz \cite{Munz} & $1.39\pm0.16$& $0.44\pm0.14$ &$0.32^{+0.16}_{-0.12}$& $24\pm3$& $5.6\pm0.6$ & $26\pm2$& $6.8\pm 1.0$\\
 Wang \cite{RSM1,RSM2} & $3.78$ & $0.501$ & $0.133$ &$48.8$ & $7.4$ & $50.3$ & $7.7$ \\
 Ebert \cite{ebert1} & $2.9$ & $0.50$ &$0.17$& $38$ & $8$ & $29$ & $6$\\
 Laverty\footnote{The values are obtained by the perturbative
(nonperturbative) calculation.} \cite{Laverty} & $2.02(2.12)$ & $0.46(0.19)$& $0.23(0.09)$ & $32.9(94.4)$ & $7.19(5.38)$ & $34.1(94.5)$& $7.59(5.57)$\\
 Schuler \cite{Schuler} & $2.5$& $0.28$ &$0.11$& $43$& $7.4$ & & \\
 Gupta\footnote{The values are obtained by
 the QCD potential (alternative treatment).} \cite{Gupta} & $6.38(8.13)$ & $0.57(1.14)$ &$0.09(0.14)$& $80(85)$ & $8(12)$ & & \\
 Huang \cite{Huang} & $3.72\pm 1.11$& $0.49\pm 0.15$& $0.13^{+0.11}_{-0.06}$&& & & \\
 Bodwin \cite{Bodwin} & $7.1\pm 2.5$ & $0.81\pm 0.29$ & $0.11\pm 0.08$&& & & \\
 Crater\footnote{The values are
 obtained by the two-body (naive) decay amplitude.} \cite{Crater} & $3.96(3.34)$& $0.743(0.435)$& $0.188(0.130)$&& & & \\
 Barbieri \cite{NR2} & $3.5$& $0.93$ & $0.27$ && & & \\
 Godfrey \cite{Godfrey} & $1.29$ & $0.46$ & $0.36$ &  &  &  & \\
 Lansberg \cite{Lansberg} & $5.00$ & $0.70$ & $0.14$ && & & \\
 Dudek\footnote{Only the statistical error is shown.} \cite{Dudek} & $2.41\pm0.58$& & && & & \\
 Lakhina \cite{Lakhina} & $3.28$ &  && & & & \\
 LO(NLO)\footnote{These values refer to a
leading (next-to-leading) order calculation done at the renormalization scale $2m_c$.} \cite{CERN} &  &  &$0.267(0.184)$& & & & \\
\end{tabular}}
\end{ruledtabular}
\end{table}
\begin{table}[ht!]
\caption{\label{tab:widthsgg} Two-gluon decay widths of the $p$-wave
heavy quarkonium states.}
\begin{ruledtabular}
\begin{tabular}{ccccccc}
  & $\Gamma^{\chi_{c0}}_{2g}$ (MeV) & $\Gamma^{\chi_{c2}}_{2g}$ (MeV) &
  $\Gamma^{\chi_{b0}}_{2g}$ (keV)& $\Gamma^{\chi_{b2}}_{2g} $ (keV)&
  $\Gamma^{\chi'_{b0}}_{2g}$ (keV)& $\Gamma^{\chi'_{b2}}_{2g}$
  (keV)\\\hline
  PDG\footnote{$\Gamma_{\rm
tot} \cong \Gamma_{2g}$.} \cite{PDG08} & $10.4\pm 0.7$ & $1.98\pm 0.11$ & & & & \\
 This work & $11.9^{+0.7}_{-0.9}$& $1.74^{-0.08}_{+0.09}$& $431^{+45}_{-49}$& $214^{-0}_{+1}$
 & $122^{+4}_{-6}$ & $92.3^{+17.7}_{-14.8}$\\
 Wang \cite{RSM1,RSM2} & $10.3$ & $2.64$ & $887$ & $220$ & $914$ & $248$ \\
 Laverty\footnote{The values are obtained by the perturbative
(nonperturbative) calculation.} \cite{Laverty} & $4.68(4.88)$ & $1.72(0.69)$ & $960(2740)$ & $330(250)$ & $990(2740)$& $350(260)$\\
 Gupta\footnote{The values are obtained by
 the QCD potential (alternative treatment).} \cite{Gupta} & $13.44(17.10)$ & $1.20(2.39)$ & $2150(2290)$ & $220(330)$& & \\
 Bodwin \cite{Bodwin} & $4.8\pm 0.7$ & $\underline{1.98\pm 0.18}$ & & & & \\
 Barbieri \cite{NR2} & $2.4$ & $0.64$ & & & & \\
 Godfrey \cite{Godfrey} & $6.25$ & $0.774$ & $672$ & $123$ & $672$ & $137$\\
 Ebert \cite{ebert2} & & & $653$ & $109$ & $431$& $76$ \\
\end{tabular}
\end{ruledtabular}
\end{table}
From these tables, except in the case of our two-photon decay width
of $\chi_{c0}$ as an input, all the $c\bar c$ decay widths of this
work are consistent with those of previous experiments and the major
theoretical methods. However, our results of $\chi_{b0,b2}$ and
$\chi'_{b0,b2}$ differ from other calculations. More specifically,
our two-photon decay widths of $\chi_{b0,b2}$ and $\chi'_{b0,b2}$
are significantly smaller than those of other approaches. From Table
2, the ratio $\Gamma(\chi_{c0}\to \gamma\gamma)/\Gamma(\chi_{b0}\to
\gamma\gamma)$ is about $300$ for our work and about $60\sim80$ for
other estimations. It is well known that the decay rate has two
contributions: one is the kinematic phase space; the other is the
dynamic decay amplitude square. No matter what approach, the phase
space and factor $e_q^4$ of $\Gamma(\chi_{c0}\to \gamma\gamma)$, in
total, is about $50$ times the one of $\Gamma(\chi_{b0}\to
\gamma\gamma)$. Regarding the comparison of decay amplitude ${\cal
M^S}$ in Eq. (\ref{Mscalar}), we find that the contribution of the
numerator in the parentheses is roughly the same for $\chi_{c0}$ and
$\chi_{b0}$. On the other hand, the denominator in the parentheses,
$M_0$, is approximately equal to the meson mass for the heavy
quarkonium. Therefore, the dynamic decay amplitude square of
$\chi_{c0}\to \gamma\gamma$ is about 6 times that of $\chi_{b0}\to
\gamma\gamma$. Our ratio $\Gamma(\chi_{c0}\to
\gamma\gamma)/\Gamma(\chi_{b0}\to \gamma\gamma)\simeq 300$ is
obtained by combining the above two components. Regardless, more
experimental measurements are required for these channels. Finally,
the ratio $R\equiv
\Gamma^{\chi_{c2}}_{2\gamma}/\Gamma^{\chi_{c0}}_{2\gamma}$ of
experiment and various theoretical estimations are also listed in
Table 2 for a comparison.

\section{Conclusions}
This study has discussed two-photon and two-gluon decay widths of
$p$-wave heavy quarkonium states within the covariant light-front
quark model. This formalism, which preserves the Lorentz covariance
in the light-front framework, was applied to annihilations of the
scalar and tensor quarkonia. To obtain the numerical results, we
used the harmonic wave functions and fixed the parameters appearing
in them. The constraints were the spin-weighted average masses
$M(S_J,~^3P_J)$ and their variational principle regarding the
Hamiltonian. We considered the linear and HO potentials in the
Hamiltonian and found that, when the data
$\Gamma(\chi_{c0}\to\gamma\gamma)$ were fitted, the former resulted
in a value of $b$ consistent with that of other quark models and the
latter led to a violation of asymptotic freedom. Therefore, only the
parameters corresponding to the linear potential were applied to
estimate the relevant decay widths. The numerical results showed
that, for the $c\bar c$ sector, all of the decay widths were in
agreement with the experimental data and the major theoretical
calculations. However, for the $b\bar b$ sector, discrepancies
appeared in the decay widths of $\chi_{b0,b2}$ and $\chi'_{b0,b2}$
from other estimations.

{\bf Acknowledgements}\\
This work is supported in part by the National Science Council of
R.O.C. under Grant No NSC-99-2112-M-017-002-MY3.

\appendix
\section{Formulas for the product of several $p_1$'s }
In general, after the $p_1$ integration, $p_1$ can be expressed in
terms of three external vectors, $\widetilde{P}$, $q$, and $\omega$,
where $\widetilde{P}=P+k$. Furthermore, the inclusion of the zero-
mode contribution cancels the $\omega$ dependence and in practice
for $p_1$ in the trace under the integration, we have
\cite{Jaus2,CCH2}
 \be
 \hat{p}_{1\mu}&\doteq& \widetilde{P}_\mu A^{(1)}_1+q_\mu A^{(1)}_2,\\
 \hat{p}_{1\mu} \hat{p}_{1\nu}&\doteq&
 g_{\mu\nu}A^{(2)}_1+\widetilde{P}_\mu \widetilde{P}_\nu A^{(2)}_2+
 (\widetilde{P}_\mu q_\nu+q_\mu \widetilde{P}_\nu)A^{(2)}_3+q_\mu q_\nu
 A^{(2)}_4+\frac{\widetilde{P}_\mu \omega_\nu+\omega_\mu
 \widetilde{P}_\nu}{\omega \widetilde{P}} B^{(2)}_1, \\
 \hat{p}_{1\mu} \hat{p}_{1\nu}\hat{p}_{1\alpha}&\doteq&
 (g_{\mu\nu}\widetilde{P}_\alpha+g_{\mu\alpha}\widetilde{P}_\nu+g_{\nu\alpha}
 \widetilde{P}_\mu)A^{(3)}_1+(g_{\mu\nu}q_\alpha+g_{\mu\alpha}
 q_\nu+g_{\nu\alpha}q_\mu)A^{(3)}_2+\widetilde{P}_\mu\widetilde{P}_
 \nu\widetilde{P}_\alpha A^{(3)}_3\non \\
 &&+(\widetilde{P}_\mu\widetilde{P}_\nu q_\alpha+\widetilde{P}_\mu q_\nu
 \widetilde{P}_\alpha+q_\mu \widetilde{P}_\nu\widetilde{P}_\alpha)
 A^{(3)}_4+(q_\mu q_\nu \widetilde{P}_\alpha+q_\mu \widetilde{P}_\nu
 q_\alpha+\widetilde{P}_\mu q_\nu q_\alpha)A^{(3)}_5\non \\
 &&+q_\mu q_\nu q_\alpha A^{(3)}_6+\frac{1}{\omega \widetilde{P}}
 (\widetilde{P}_\mu\widetilde{P}_\nu \omega_\alpha+\widetilde{P}_\mu
 \omega_\nu \widetilde{P}_\alpha+\omega_\mu \widetilde{P}_\nu
 \widetilde{P}_\alpha)B^{(3)}_1\non \\
 &&+\frac{1}{\omega \widetilde{P}}
 [(\widetilde{P}_\mu q_\nu+q_\mu \widetilde{P}_\nu) \omega_\alpha+
 (\widetilde{P}_\mu q_\alpha+q_\mu \widetilde{P}_\alpha)
 \omega_\nu (\widetilde{P}_\nu q_\alpha+q_\nu
 \widetilde{P}_\alpha)\omega_\mu]B^{(3)}_2,\\
 \hat{p}_{1\mu} \hat{p}_{1\nu}\hat{p}_{1\alpha}\hat{p}_{1\beta}&\doteq&
 \sum^9_{i=1} I_{i\mu\nu\alpha\beta} A^{(4)}_i+\sum^4_{j=1}
 J_{j\mu\nu\alpha\beta} B^{(4)}_j,
 \en
where the symbol $\doteq$ denotes that the equation is true only in
the integration and
 \be
 I_{1\mu\nu\alpha\beta}&=&g_{\mu\nu}g_{\alpha\beta}+g_{\mu\alpha}
 g_{\nu\beta}+g_{\mu\beta}g_{\nu\alpha},\non \\
 I_{2\mu\nu\alpha\beta}&=&g_{\mu\nu}\widetilde{P}_{\alpha}\widetilde{P}_{\beta}
 +g_{\mu\alpha}\widetilde{P}_{\nu}\widetilde{P}_{\beta}+g_{\mu\beta}
 \widetilde{P}_{\nu}\widetilde{P}_{\alpha}+g_{\alpha\beta}
 \widetilde{P}_{\mu}\widetilde{P}_{\nu}+g_{\nu\beta}\widetilde{P}_{\mu}
 \widetilde{P}_{\alpha}+g_{\nu\alpha}\widetilde{P}_{\mu}\widetilde{P}_{\beta},\non \\
 I_{3\mu\nu\alpha\beta}&=&g_{\mu\nu}(\widetilde{P}_{\alpha}q_\beta+
 \widetilde{P}_{\beta}q_\alpha)+{\rm permutations},\non \\
 I_{4\mu\nu\alpha\beta}&=&g_{\mu\nu}q_{\alpha}q_{\beta} +g_{\mu\alpha}q_{\nu}q_{\beta}
 +g_{\mu\beta}q_{\nu}q_{\alpha}+g_{\alpha\beta}q_{\mu}q_{\nu}
 +g_{\nu\beta}q_{\mu}q_{\alpha}+g_{\nu\alpha}q_{\mu}q_{\beta},\non \\
 I_{5\mu\nu\alpha\beta}&=&\widetilde{P}_{\mu}\widetilde{P}_{\nu}
 \widetilde{P}_{\alpha}\widetilde{P}_{\beta},\non \\
 I_{6\mu\nu\alpha\beta}&=&\widetilde{P}_{\mu}\widetilde{P}_{\nu}
 \widetilde{P}_{\alpha}q_{\beta}+\widetilde{P}_{\mu}\widetilde{P}_{\nu}
 q_{\alpha}\widetilde{P}_{\beta}+\widetilde{P}_{\mu}q_{\nu}
 \widetilde{P}_{\alpha}\widetilde{P}_{\beta}+q_{\mu}\widetilde{P}_{\nu}
 \widetilde{P}_{\alpha}\widetilde{P}_{\beta},\non \\
 I_{7\mu\nu\alpha\beta}&=&\widetilde{P}_{\mu}\widetilde{P}_{\nu}
 q_{\alpha}q_{\beta}+{\rm permutations},\non \\
 I_{8\mu\nu\alpha\beta}&=&\widetilde{P}_{\mu}q_{\nu}
 q_{\alpha}q_{\beta}+q_{\mu}\widetilde{P}_{\nu}
 q_{\alpha}q_{\beta}+q_{\mu}q_{\nu}\widetilde{P}_{\alpha}q_{\beta}+q_{\mu}q_{\nu}
 q_{\alpha}\widetilde{P}_{\beta},\non \\
 I_{9\mu\nu\alpha\beta}&=&q_{\mu}q_{\nu}q_{\alpha}q_{\beta},\non \\
 J_{1\mu\nu\alpha\beta}&=&\frac{1}{\omega\widetilde{P}}[g_{\mu\nu}(\widetilde{P}_{\alpha}
 \omega_{\beta}+\widetilde{P}_{\beta}\omega_\alpha)+{\rm permutations}],\non \\
 J_{2\mu\nu\alpha\beta}&=&\frac{1}{\omega\widetilde{P}}(\widetilde{P}_{\mu}\widetilde{P}_{\nu}
 \widetilde{P}_{\alpha}\omega_{\beta}+\widetilde{P}_{\mu}\widetilde{P}_{\nu}
 \omega_{\alpha}\widetilde{P}_{\beta}+\widetilde{P}_{\mu}\omega_{\nu}
 \widetilde{P}_{\alpha}\widetilde{P}_{\beta}+\omega_{\mu}\widetilde{P}_{\nu}
 \widetilde{P}_{\alpha}\widetilde{P}_{\beta}),\non \\
 J_{3\mu\nu\alpha\beta}&=&\frac{1}{\omega\widetilde{P}}[(\widetilde{P}_{\mu}\widetilde{P}_{\nu}
 q_{\alpha}+\widetilde{P}_{\mu}q_{\nu}\widetilde{P}_{\alpha}+q_{\mu}\widetilde{P}_{\nu}
 \widetilde{P}_{\alpha}) \omega_\beta+{\rm permutations}],\non \\
 J_{4\mu\nu\alpha\beta}&=&\frac{1}{\omega\widetilde{P}}[(\widetilde{P}_{\mu}q_{\nu}
 q_{\alpha}+q_{\mu}\widetilde{P}_{\nu}q_{\alpha}+q_{\mu}q_{\nu}
 \widetilde{P}_{\alpha}) \omega_\beta+{\rm permutations}],
 \en
with
 \be
 &&A^{(1)}_1=\frac{x_1}{2},\quad
 A^{(1)}_2=A^{(1)}_1-\frac{p_\bot\cdot q_\bot}{q^2},\quad A^{(2)}_1=-p^2_\bot
 -\frac{(p_\bot\cdot q_\bot)^2}{q^2},\quad A^{(2)}_2=(A^{(1)}_1)^2,\non \\
 && A^{(2)}_3=A^{(1)}_1 A^{(1)}_2,\quad A^{(2)}_4=(A^{(1)}_2)^2
 -\frac{A^{(2)}_1}{q^2},\quad B^{(2)}_1=A^{(1)}_1 Z_2-A^{(2)}_1,\quad A^{(3)}_1=A^{(1)}_1
 A^{(2)}_1,\non \\
 &&A^{(3)}_2=A^{(1)}_2 A^{(2)}_1,\quad A^{(3)}_3=A^{(1)}_1
 A^{(2)}_2,\quad A^{(3)}_4=A^{(1)}_2 A^{(2)}_2,\quad
 A^{(3)}_5=A^{(1)}_1 A^{(2)}_4,\non \\
 &&A^{(3)}_6=A^{(1)}_2 A^{(2)}_4-\frac{2}{q^2} A^{(1)}_2
 A^{(2)}_1,\quad B^{(3)}_1=A^{(1)}_1(B^{(2)}_1-A^{(2)}_1),\quad
 B^{(3)}_2=A^{(1)}_2 B^{(2)}_1+\frac{q\cdot \widetilde{P}}{q^2}
 A^{(3)}_1,\non \\
 &&A^{(4)}_1=\frac{(A^{(2)}_1)^2}{3},\quad A^{(4)}_2=A^{(1)}_1
 A^{(3)}_1,\quad A^{(4)}_3=A^{(1)}_1 A^{(3)}_2,\quad
 A^{(4)}_4=A^{(1)}_2A^{(3)}_2-\frac{A^{(4)}_1}{q^2},\non \\
 &&A^{(4)}_5=A^{(1)}_1 A^{(3)}_3,\quad A^{(4)}_6=A^{(1)}_1
 A^{(3)}_4,\quad A^{(4)}_7=A^{(1)}_1 A^{(3)}_5,\quad
 A^{(4)}_8=A^{(1)}_1 A^{(3)}_6,\non \\
 &&A^{(4)}_9=A^{(1)}_1 A^{(3)}_6-\frac{3}{q^2} A^{(4)}_4,\quad
 B^{(4)}_1=A^{(1)}_1 A^{(2)}_1 Z_2-A^{(4)}_1,\quad
 B^{(4)}_2=A^{(1)}_1 B^{(3)}_1-A^{(4)}_2,\non \\
 &&B^{(4)}_3=A^{(1)}_1 B^{(3)}_2-A^{(4)}_3,\quad
 B^{(4)}_4=A^{(1)}_1\left(A^{(2)}_4 Z_2+2\frac{q\cdot \widetilde{P}}{q^2}
 A^{(1)}_2 A^{(2)}_1\right)-A^{(4)}_4,\non \\
 && Z_2=\hat{N}_1+(x_2-x_1) M^2+(q^2+q\cdot \widetilde{P})\frac{p_\bot \cdot q_\bot}{q^2}. \label{A6}
 \en
Equation (\ref{A6}) is obtained by contracting the larger number of
$\hat{p}_1$'s with $\omega_\beta$, $q_\beta$, and $g_{\alpha\beta}$,
and through a comparison with the complete expression of the fewer
$\hat{p}_1$'s.


\end{document}